\documentclass[aps, prl, twocolumn, reprint,preprintnumbers, nofootinbib, showpacs, superscriptaddress]{revtex4}%draft
\usepackage{graphicx, amsmath, amssymb, amsfonts, pifont, bm, cancel}
\usepackage[usenames]{color}
\usepackage{subfigure}
\usepackage{braket}
\usepackage[normalem]{ulem} % either use this (simple) or

\definecolor{MyDarkBlue}{rgb}{0,0,1}

\newcommand{\be}[0]{\begin{equation}}
\newcommand{\ee}[0]{\end{equation}}
\newcommand{\bea}[0]{\begin{eqnarray}}
\newcommand{\eea}[0]{\end{eqnarray}}

 \newcommand{\beq}{\begin{equation}}
 \newcommand{\eeq}{\end{equation}}
 \newcommand{\bel}{\begin{align*}}
 \newcommand{\tamam}{\end{align*}}

 \newcommand{\beqa}{\begin{eqnarray}}             %Begin Equation Array
 \newcommand{\eeqa}{\end{eqnarray}}               %End Equation Array
 \newcommand{\ipr}[2]{\left\langle#1|#2\right\rangle}

\begin{document}
% The following information is for internal review, please remove them for submission
\widetext
%\leftline{Version xx as of \today}
%\leftline{To be submitted to PRL}
\leftline{Comment to {\tt mirhosse@optics.rochester.edu}}
 %\preprint{APS/123-QED}
\title{Compressive direct measurement of the quantum wavefunction}

\author{M.~Mirhosseini}
\email{mirhosse@optics.rochester.edu}
%\thanks{To whom correspondence should be addressed. E-mail: zshi@optics.rochester.edu}
%\noaffiliation
%\affiliation{The Institute of Optics, University of Rochester, Rochester, New York 14627, USA}
\author{O.~S.~Maga\~na-Loaiza}
\affiliation{The Institute of Optics, University of Rochester, Rochester, New York 14627, USA}
\author{S. M. Hashemi Rafsanjani}
\affiliation{Center for Coherence and Quantum Optics and the Department of Physics \& Astronomy,University of Rochester, Rochester, New York 14627, USA}
\author{R.~W.~Boyd}
\affiliation{The Institute of Optics, University of Rochester, Rochester, New York 14627, USA}
\affiliation{Department of Physics, University of Ottawa, Ottawa ON K1N 6N5, Canada}

\date{\today}

\begin{abstract}
The direct measurement of a complex wavefunction has been recently realized by using weak-values. In this paper, we introduce a method that exploits sparsity for compressive measurement of the transverse spatial wavefunction of photons. The procedure involves a weak measurement in random projection operators in the spatial domain followed by a post-selection in the momentum basis. Using this method, we experimentally measure a 192-dimensional state with a fidelity of $90\%$ using only $25$ percent of the total required measurements. Furthermore, we demonstrate measurement of a 19200 dimensional state; a task that would require an unfeasibly large acquiring time with the conventional direct measurement technique.
\end{abstract}

\pacs{42.30.Ms, 42.50.Ar, 42.30.Va}
\maketitle

%\section{Introduction}
The no-clonning theorem prohibits exact determination of the quantum wavefunction from a single measurement \cite{Wootters:1982ex,Dieks:1982gj,Milonni:1982es}. In contrast, a large ensemble of identically prepared quanta can be used to estimate the wavefunction through quantum state tomography. This procedure is well-known and has been implemented in different scenarios \cite{Kanseri:2012tb,Cramer:2010bs,Hofheinz:2009ba,Resch:2005cc,Beck:2001dr,James:2001bb,Smithey:1993er,Vogel:1989dy}. However, tomography involves a time-consuming computationally complex post-processing, and its implementation becomes inevitably more challenging as the dimension of the Hilbert space increases \cite{James:2001bb, Agnew:2011js}. Due to the difficulty of state determination in such high dimensional systems, efficient measurement methods for characterizing pure and mixed states are desirable. 

%Measurement is very challenging in quantum mechanics. Measurement of a wavefunction of a single particle is impossible due to no-cloning theorem. It is however possible to estimate the wavefunction if many identical realizations are available. This procedure is usually time-consuming and computationally complex. Due to this, efficient measurement methods for characterizing pure and mixed states are desirable. 

Recently, there has been tremendous interest in determining the complex wavefunction of a pure state through the use of weak-values \cite{Lundeen2011,Salvail:2013bo,Malik:2014bf}. This method, known as the direct measurement method, provides a convenient procedure for estimation of a wavefunction. It has been suggested the direct measurement (DM) is an efficient means for characterizing high-dimensional states due to the simplicity of realization and absence of a time-consuming post processing \cite{Lundeen2011}. Yet, the measurement of high-dimensional states remains a challenging task. Even for DM the number of measurements that are needed to characterize the state vectors grows linearly with the dimension of the state. Further, a much larger ensemble of identically prepared particles is required for reliable measurement of elements of the state vector in a high-dimensional Hilbert space \cite{Maccone:2014if}.

In this Letter, we introduce a method which combines the benefits of direct measurement with a novel computational technique known as compressive sensing \cite{Shabani:2011de, Howland:2013gi,Liu:2012im,Gross:2010cv,Katz:2009fv,Baraniuk:2008vn}. Utilizing our approach, the wavefunction of a high-dimensional state can be estimated with a high fidelity using much fewer number of measurements than a simple direct measurement approach. In the following we first briefly discuss the direct measurement and then propose compressive direct measurement (CDM). We then describe our experimental implementation of CDM, which provides a direct test of this method. In our experiment, we were able to reconstruct a wavefunction with only a fraction of the required measurements for a DM measurement with a more than 90 percent fidelity. 

We explain and implement DM and CDM for the case of a transverse photonic state. Thus we closely follow the experimental setup that was originally implemented in \cite{Lundeen2011}. Yet the mathematics and ideas can be generalized for other quantum wavefunctions. Note that in practice the transition from the continuous spatial domain to a discrete state vector can be achieved by dividing the continuous coordinate to a finite number of pixels. In this case the coefficient for each element of the discrete state vector equals the value of the corresponding continuous wavefunction averaged over a small pixel area. Hence, the pixel sizes should be chosen sufficiently small to include all the features of the specific group of wavefunctions of interest. 

A weak value is the expectation value of a weak measurement that is followed by a post-selection \cite{Aharonov:1988fk}. Now consider a weak measurement of the position projector $\hat{\pi}_j = \ket {x_j} \bra {x_j}$ at point $x_j$ followed by a post-selection on the zeroth component of the Fourier transform of the spatial wavefunction, which we denote by $\ket{o}$. The expectation value of the pointer state after post-selection in this case can be calculated using the weak-value formula
 \begin{align}\label{DM}
 \pi_{w} = \frac{\braket{o | x_j}\braket{x_j | \psi}}{\braket{o | \psi}} = \frac{\psi(x_j) }{{\phi_{0}\sqrt{N}}}
 \end{align}
where $\phi_{0}=\ipr{o}{\psi}$. We have used the Fourier transform property $\ipr{o}{x_{j}}=1/\sqrt{N}$ where $N$ is the dimension of the Hilbert space. We treat $\phi_{0}$ as a real number. This leads to no loss of generality since the wavefunction can always be multiplied by a factor with appropriate phase to achieve this condition. 
 Consequently, the complex wavefunction can be calculated at each point by measuring the real and imaginary part of the weak value $ \pi_{w}$.
%\begin{align}\label{DM}
%&\psi_j=\frac{\phi_{0}\sqrt{d}}{2 \alpha} [{\bar{\sigma}_{y,j}+i \bar{\sigma}_{x,j}}].
%\end{align}

%\section{Compressive direct measurement} We denote the set of $Q_{m,j}$ for all the values of $j$ as the sensing vector $\mathbf{Q}$.
We now generalize the DM to a form suitable for compressive sensing. Let the initial system-pointer state be
\begin{align}
\ket{\Omega}=\ket{\psi}\otimes\ket{V}=\sum_{i=1}^N \psi_i\ket{x_i}\otimes\ket{V}.
\end{align}
where we have assumed to have a discrete Hilbert space for the spatial degree of freedom $\ket{\psi}$ and a two-level system such as the polarization of a single photon for the pointer state $\ket{V}$. We consider a situation where instead of a measuring a projector $\hat{\pi}_j$ we perform a weak measurement of the operator  $\hat{Q}_m=\sum_j Q_{m,j} \hat{\pi}_j$ where the coefficients $Q_{m,j} \in \mathbb{R}$:
%\begin{align}
%\hat{Q}_m=\sum_j Q_{m,j} \hat{\pi}_j ,
%\end{align}
The effect of this measurement can be described by making a Taylor series approximation to the measurement's evolution operator $\exp(-i \alpha~\hat{\pi_{i}}\otimes \hat{\sigma}_{x})$.  Here, $\hat{\sigma}_{x}$ is a Pauli matrix and $\alpha$ is the angle of rotation of the polarization.
\begin{align}
e^{-i\alpha \hat{Q}_m\otimes \hat{\sigma}_{x}}\ket{\Omega} \approx   
   \ket{\Omega}-i\alpha\sum_j Q_{m,j} ~\psi_{j} \ket{x_j}\otimes\ket{H}.     
\label{expansion}
\end{align}
Now we consider post-selection on $\ket{o}$. In this case, we are left with a polarization state with no spatial degree of freedom (Note that $\ket{o}$ is not the vacuum state of the electric field). A weak measurement of the operator $\hat{Q}_m$ followed by a post-selection on $\ket{o}$ leads to
\begin{align}\label{pol-final}
\ket{s_m}=\ket{V}-i \frac{\alpha}{\phi_{0}~\sqrt{N}} \sum_j Q_{m,j}{ \psi_j} \ket{H}
\end{align}

Note that physically, the weak measurement of operator $\hat{Q}_m$ is equivalent to a rotation of polarization at each point $x_j$ by the value $\alpha~Q_{m,j}$. In this situation the expected values of the polarization of the post-selected state can be written as

\begin{align}
&\bar{\sigma}_{x,m}\equiv\bra{s_m}\hat{\sigma}_{x}\ket{s_m}=\kappa \sum_j Q_{m,j} \Im[\psi_j],\\
&\bar{\sigma}_{y,m}\equiv\bra{s_m}\hat{\sigma}_{y}\ket{s_m}=\kappa \sum_j Q_{m,j} \Re[\psi_j],
\end{align}
where $\Im[\psi_{j}]$ and $\Re[\psi_{j}]$ are the imaginary and the real part of $\psi_{j}$ respectively and $\kappa = \frac{2\alpha}{\phi_{0} \sqrt{N}}$. Combining the results $\bar{\sigma}_{x,m}$ and $\bar{\sigma}_{y,m}$  to a complex value $ \phi_m = 1/\kappa [\bar{\sigma}_{x,m}+ i \bar{\sigma}_{y,m}]$ and repeating the measurement several times we a set of linear equations

\begin{equation}
 \begin{pmatrix}
   \phi_1 \\
   \phi_2 \\
   \vdots \\
   \phi_{M}
\end{pmatrix}=
 \begin{pmatrix}
   Q_{1,1} & Q_{1,2} & \cdots & Q_{1,N} \\
   Q_{2,1} & Q_{2,2} & \cdots & Q_{2,N} \\
   \vdots & \vdots & \ddots & \vdots \\
    Q_{M,1} & Q_{M,2} & \cdots & Q_{M,N}
\end{pmatrix}
 \begin{pmatrix}
   \psi_1 \\
   \psi_2 \\
   \vdots \\
   \psi_{N}
\end{pmatrix}.
\end{equation}

Writing the equations above in a more compact form we have
\begin{equation}
\boldsymbol{\phi}=\mathbf{Q}~\boldsymbol{\psi}.
\end{equation}

Here, $ m\in \{1:M\}$ and $ n\in \{1:N\}$ where $M$ is the total number of sensing operators and $N$ is the dimension of the Hilbert state of the unknown wavefunction. To find the wavefunction $\psi$ we need to solve the above linear system of equations. For the special case $M =N$ the set of equations can be exactly solved for a non-singular matrix $\mathbf{Q}$. However, we are interested in the case where $M\leq N$. The pseudo-inverse of $\mathbf{Q}$ can be used as an optimal linear recovery strategy to find a solution that minimizes the least square error:
\begin{equation}\label{LinearConst}
\boldsymbol{\psi}=\mathbf{Q}^\dagger (\mathbf{Q}\mathbf{Q}^\dagger)^{-1}\boldsymbol{\phi}.
\end{equation}
 However, a nonlinear strategy can be used to recover $\boldsymbol{\psi}$ with a far superior quality using the idea of compressive sensing (CS). Consider a linear transformation represented by matrix $\mathbf{T}$. If the wavefunction under the experiment $\boldsymbol{\phi}$ is known to have very few non-zero coefficients under this transformation, $\boldsymbol{\psi}$ can be recovered by solving the convex optimization problem \cite{romberg:2008gq}
\begin{equation}\label{CS}
\min_{\boldsymbol{\psi'}} ||{\mathbf{T}\boldsymbol{\psi'} }||_{\ell_1}\text{ , subject to }\mathbf{Q} \boldsymbol{\psi'} = \boldsymbol{\phi}.
\end{equation}

\begin{figure}
\begin{center}
\includegraphics[width=1\columnwidth]{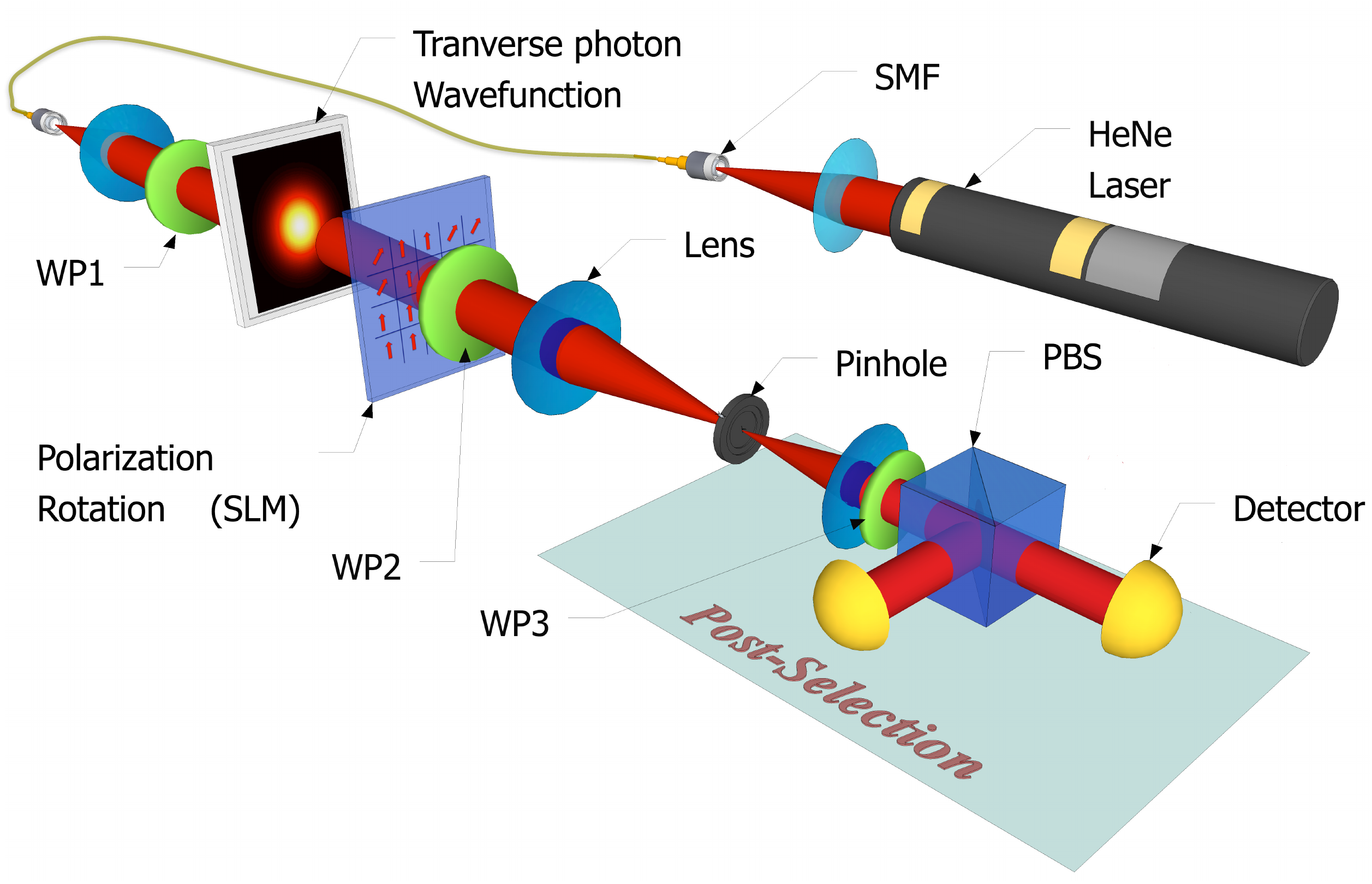}
\caption{A schematic illustration of the experimental setup.}
\label{schematic}
\end{center}
\end{figure}

\begin{figure}[]
\begin{center}
\includegraphics[width=1\columnwidth]{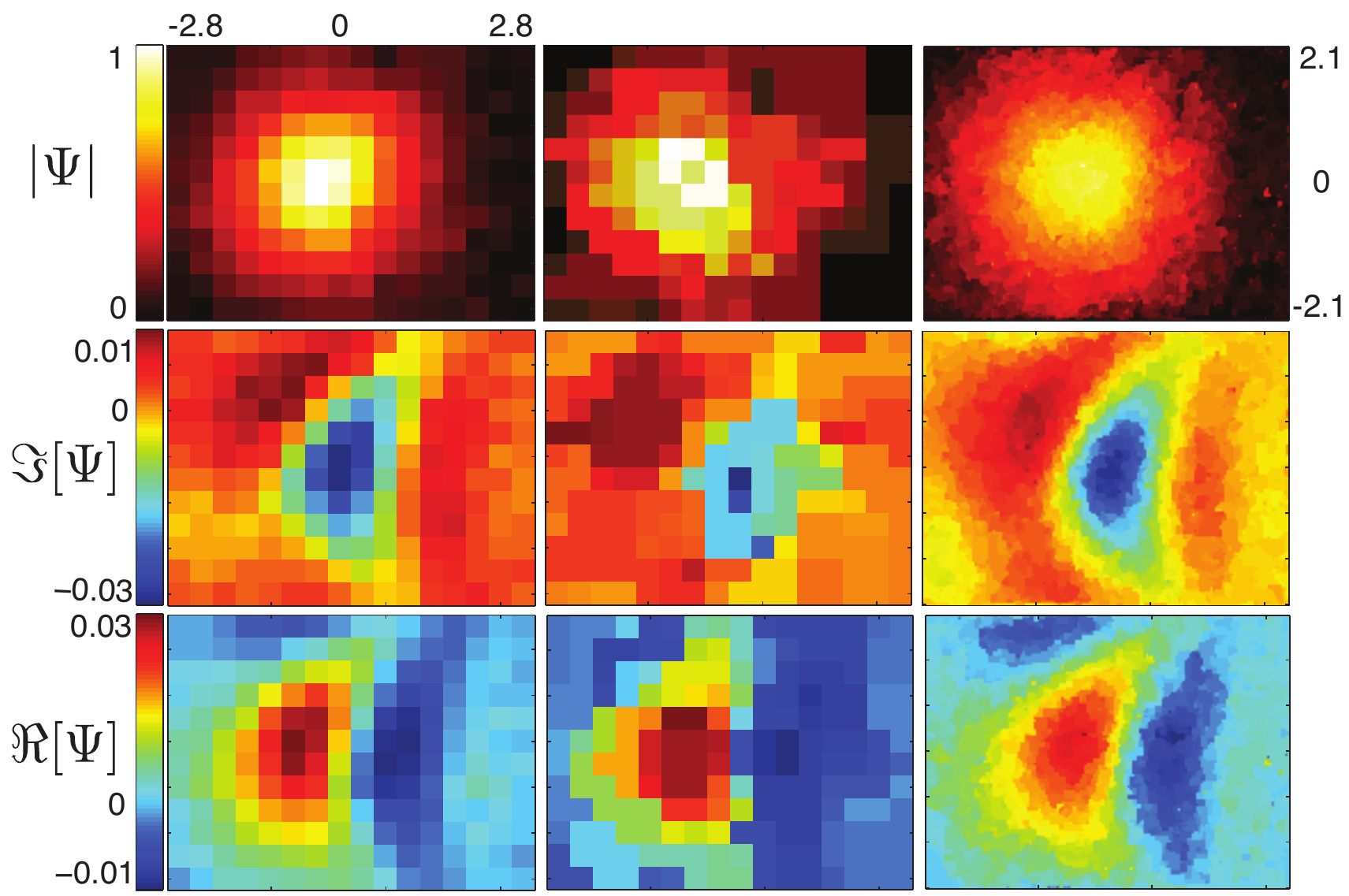}
\caption{The amplitude, real, and imaginary parts of an aberrated Gaussian state from experimental data. The left column presents data from a pixel-by-pixel scan of the state for N=$192$. The middle column shows the reconstructed wavefront for for N=$192$, and $M/N = 20\%$ of total measurements from the CDM method. The right column demonstrates reconstruction N=$19200$, and $M/N = 20\%$ of total measurements. The transverse dimensions of the state are shown in milimiters.}
\label{Pics}
\end{center}
\end{figure}

\noindent where $||\cdot||_{\ell1}$ represents the $1$-norm. For this approach to work, it is critical that the two bases, defined by $Q$ and $T$, are incoherent \cite{romberg:2008gq}. The coherence of the two bases is defined by the square root of the dimension of the bases times the highest fidelity between any pairs of states from the two bases \cite{Candes:2007vv}. According to CS theory if the coherence of the two bases is small, by an overwhelming probability, the target wavefunction $\boldsymbol{\psi}$ can be recovered with $M\geq O[K\log(N)]$ measurements, where $K$ is the number of nonzero components of $\mathbf{T}\boldsymbol{\psi}$ \cite{Candes:2007vv}. Functions with spatial correlations are shown to be extremely likely to have sparse coefficients in discrete cosine transform or wavelet transform domains \cite{romberg:2008gq, Zerom:2011fu}. However, a much simpler variant of Eq. (\ref{CS}) can be used in practice to achieve results of comparable quality \cite{romberg:2008gq,MaganaLoaiza:2013ut}. In this method the target wavefunction can be found by optimizing the quantity

\begin{equation}\label{CS2}
\min_{\boldsymbol{\psi'}} \sum_j {||{\nabla\boldsymbol{\psi'_j} }||}_{\ell_1}+\frac{\mu}{2}{||\mathbf{Q} \boldsymbol{\psi'} - \boldsymbol{\phi}||}_{\ell_2}^2. 
\end{equation}
Here, ${\nabla\boldsymbol{\psi'_j} }$ is the discrete gradient of $\boldsymbol{\psi'}$ at position $x_j$ and $\mu$ is a penalty factor. Heuristically, the minimization of the first term results in a smooth function while the second factor minimizes deviations from the experimental results $\boldsymbol{\phi}$. The optimal value of $\mu$ should be chosen considering the specifics of the target wavefunction and the signal-to-noise ratio of the experimental data. At the end we retrieve the wavefunction from the solution of the optimization problem as $\ket{\psi'} = \sum_{i=1}^N \psi'_i\ket{x_i}$.

%, $\boldsymbol{\psi'}$,In practice, it is desirable to restrict the possible values of  $Q_{m,j}$ to few choices. A common example for $Q_{m,j} \in \{1,-1\}$ are elements of a Hadamard matrix. Due to defining properties of Hadamard matrices such sensing vectors $Q_{m}$ form an orthogonal basis set. Alternatively, each vector $Q_{m}$ can be created by randomly choosing the elements $Q_{m,j}$ from the set $\{0,1\}$ with equal probabilities.
\begin{figure}[t]
\centerline{\includegraphics[width=\columnwidth]{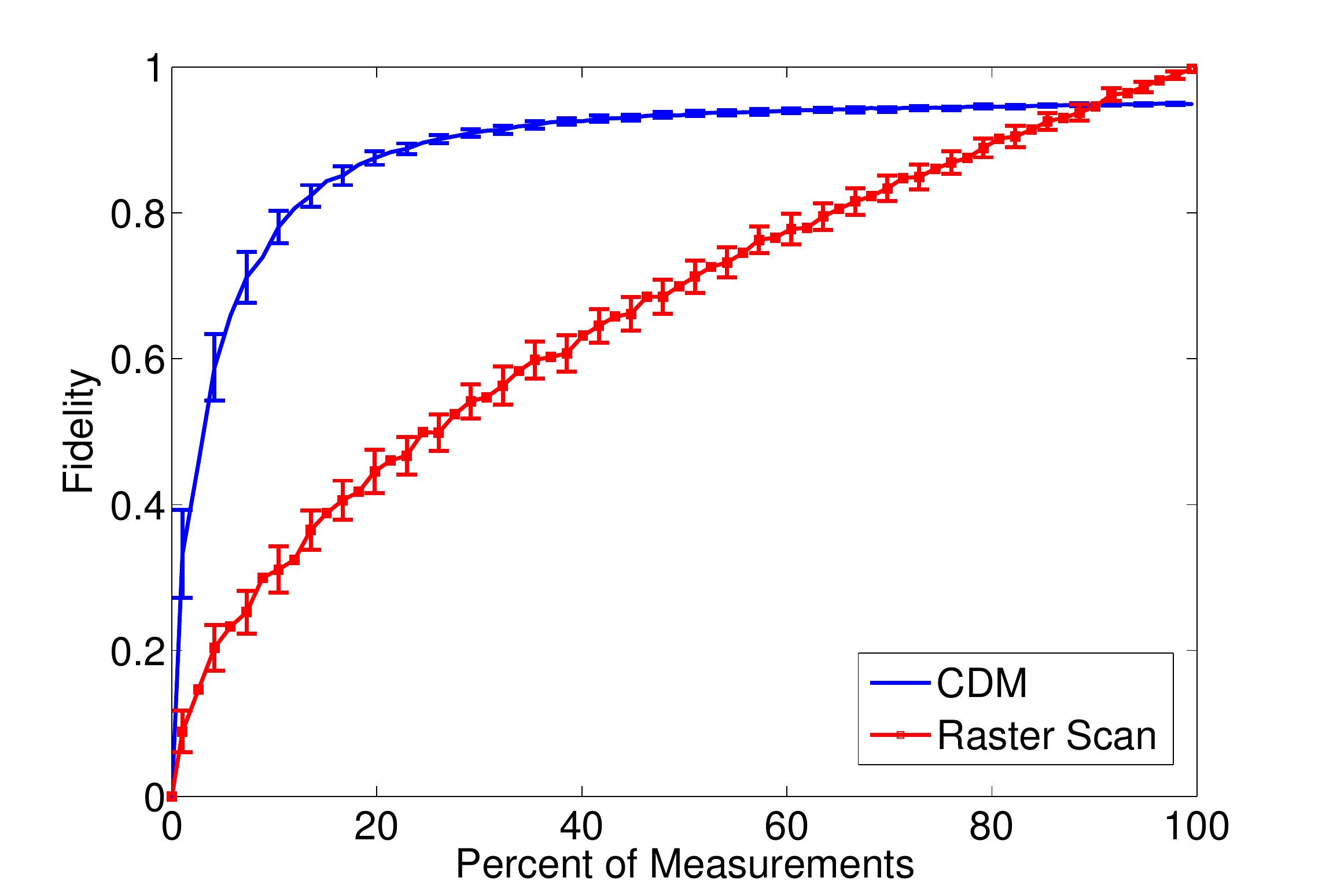}}
\caption{The fidelity of the reconstructed state with the target wavefunction as a function of the percentage of the total measurements. The fidelity of the state reconstructed with CDM is shown in blue. The fidelity of the state reconstructed from a partial pixel-by-pixel scan with the same number of measurements is shown in red for comparison. The error-bars represent standard deviation calculated from 100 repetitions of the experiment (error-bars are shown at every second data point for visual clarity).}
\label{Fidelity}
\end{figure}

%\section{Experiment}

Fig.~\ref{schematic} shows the schematics of the experiment. A vertically polarized Gaussian mode is prepared by spatially filtering a He-Ne laser beam with a single mode fiber and passing it through a polarizer. The polarization rotation is performed using a spatial light modulator (SLM) in combination with two quarter wave plates (QWP) \cite{mirhosseini2012photon,Moreno2007}. The SLM provides the ability to rotate the polarization of the incident beam at every single pixel in a controlled fashion. The post-selection in the momentum basis is done using a Fourier-transforming lens and a single mode pinhole. We retrieve the real part of the weak value using a combination of a half wave plate (HWP) and a polarizing beam splitter (PBS). The beams from the output ports of the beam splitter are coupled to single mode fibers that are connected to avalanche photo-diodes (APDs). Similarly, the imaginary part of the weak value is measured by replacing the  HWP (shown as WP3) with a QWP. 

 We perform a random polarization rotation of either $\alpha = 20^{\circ}$ or zero at each pixel, corresponding to $Q_{m,j}$ values of 1 and 0. For different values of $m$, we load different pre-generated sensing vectors $\mathbf{Q}_m$ onto the SLM and repeat the experiment. The wavefunction is then retrieved via post processing on a computer. We use the algorithm known as ÒTotal Variation Minimization by Augmented Lagrangian and Alternating DirectionÓ (TVAL3) \cite{Li:2009um} to solve Eq.\,(\ref{CS2}). Our target wavefunction is the Gaussian mode mode from the fiber single onto the SLM. The lens after the fiber is slightly displaced to create an aberrated wavefront. This create a complex wavefunction made from both real and imaginary parts.

We reconstruct the wavefunction from the conventional direct measurement method using Eq.\,(\ref{DM}). The real and imaginary parts from a pixel-by-pixel raster scan are shown on the left column of Fig.\,\ref{Pics} for a $N = 12 \times 16 = 192$ dimensional Hilbert space. The real and imaginary parts of the wavefunction reconstructed from CDM using $M \approx 0.2 N = 38 $ are shown on the middle column. It can be seen that the main features of the state are retrieved with as few as $20\%$ of the total number of measurements used in the left column. It should be emphasized that the minimum number of required measurement for an accurate reconstruction is proportional to the sparsity of the signal. Our algorithm uses sparsity with respect to the gradient transformation, according to Eq. \ref{CS2}. In order to achieve a more sparse signal, we have done a fine grain measurement of the same state at the resolution of $N = 120 \times 160 = 19200$. The wavefunction reconstructed from CDM using $M \approx 0.2 N = 3800 $ is shown on the right column of Fig.\,\ref{Pics}. Due to increased sparsity of the state in the larger Hilbert space, a very detailed reconstruction can be achieved with $20\%$ of the total number of measurements.
 
To provide a quantitive comparison of the two methods we calculate the fidelity between the retrieved state $\ket{\psi'}$ and the target state $\ket{\psi}$ from a full pixel-by-pixel scan as
\begin{equation}
F(\ket{\psi'},\ket{\psi}) = |\braket{\psi'|\psi}|.
\end{equation}

The results are shown in Fig.\,\ref{Fidelity}. The horizontal axis corresponds to the percentage of the measurements ($100 \times \frac{M}{N}$). The blue curve shows the fidelity of the state reconstructed with the CDM method. The red curve represents the average fidelity of state reconstructed with Eq.\,(\ref{LinearConst}) using the data from a partial pixel-by-pixel measurement of $M$ randomly chosen points. It is seen from the figure that the compressive method results in a drastic increase of fidelity for the first few measurement and gradually settles to a value close to $1$. As an example of the usefulness of the compressive method, a fidelity as high as $90\%$ is achieved by performing only $25\%$ of measurements, while the conventional direct measurement needs approximately $80\%$ of all the measurements to achieve the same value of fidelity.

\begin{figure}[]
\begin{center}
\includegraphics[width=\columnwidth]{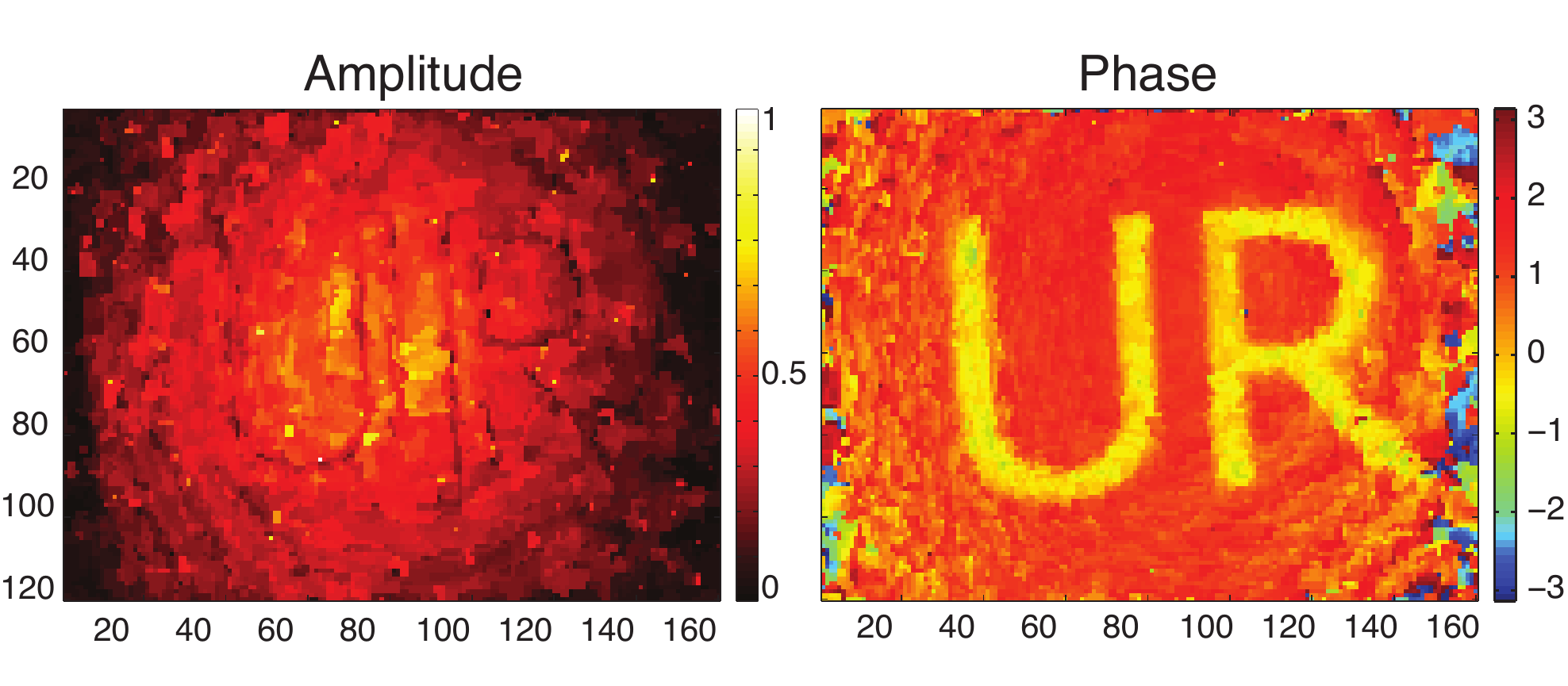}
\caption{The amplitude and phase of a Gaussian mode illuminating a custom phase mask (the logo of the University of Rochester). The data is reconstructed by the CDM method with N=$19200$, and $M/N = 20\%$ of total measurements. }
\label{UR}
\end{center}
\end{figure}

To further demonstrate the accuracy of our method we have used it to measure a custom state prepared using a phase mask depicting letters U and R with a phase jump of $\pi/2$. The phase mask is prepared via an additional spatial light modulator illuminated with the Gaussian beam from the laser and the state is imaged onto the second SLM which is used for polarization rotation. Figure \ref{UR} shows the amplitude and the phase of the reconstructed state with $M/N = 20\%$ of the total measurements. Notice that while the amplitude is relatively uniform, the phase shows the letters U and R with a remarkable accuracy. It should be emphasized that the measurement of a state of such high dimensions is extremely time consuming via a pixel-by-pixel scanning. In our approach, we perform a weak measurement on approximately half of all the pixels at each time. Due to this, the change in the state of the pointer (i.e. the polarization of the beam after the pinhole) is much more pronounced as compared to the conventional DM where only one pixel would be weakly measured. The speed-up factor can be estimated considering that the strength of the signal measured in the laboratory is proportional to the value of the second term in Eq. \ref{pol-final}. It is easy to check that the magnitude of this term is on average $\sqrt{N}/2$ larger in the case when half of $Q_{m,j}$ are set to one. For the case of our experiment with $N = 19200$, and $M/N = 20\%$, our approach provides a $\sim350$-fold speed-up in the measurement procedure.

It should be emphasized that our specific experimental realization of the CDM method can be described using classical physics. The measured wavefunction in this case is the spatial mode of photons which is equivalent to the electric field of paraxial light beams in the classical limit \cite{Bamber2012a}. Since the experiment is designed to measure the spatial mode, it is insensitive to the number of excitations of the field (i.e. the number of photons). Subsequently, the results of the experiment would be the same for a source of single photons, heralded single photons or a strong laser beam provided that they are prepared in the same spatial and polarization modes. However, the language of quantum mechanics provides a simpler description, with a broader range of applicability that includes fundamentally quantum mechanical states such as electron beams.

%\section{Summary}

Determining an unknown wavefunction is of fundamental importance in quantum mechanics. Despite many seminal contributions, in practice this task remains challenging, especially for high-dimensional states. The direct measurement approach, introduced by Lundeen \emph{et.~al}, has provided a ground for meeting the high-dimensionality challenge \cite{Lundeen2011}. Here we combine the efficiency of compressive sensing with the simplicity of the direct measurement in determining the wavefunction of an a priori unknown state. Our experimental results demonstrate that a compressive variation of the direct measurement allows an accurate determination of a 192-dimensional state with a fidelity of $90\%$ using only $25$ percent of measurements that are needed for a simple direct measurement approach. This method provides an easy means of characterizing high-dimensional systems in the labs. In addition, the technique can be used for classical applications which involve a classical beam of light such as wavefront sensing.

We acknowledge helpful discussions with J. H. Eberly, B. Rodenburg and Z. Shi.

\bibliography{CDMRef}{}
\end{document}